\documentclass[a4paper,twocolumn]{esapub2005} 

\usepackage{times}
\usepackage{natbib}
\usepackage{graphicx}

\begin{document}

\pagestyle{empty}

\title{NEW OBSERVATIONS AND MODELS OF THE KINEMATICS OF THE ZODIACAL DUST CLOUD}
\author{G.J. Madsen}
\affil{Anglo-Australian Observatory, P.O. Box 296, Epping, NSW 1710, Australia, Email: madsen@aao.gov.au}
\author{R.J. Reynolds}
\affil{Univ. of Wisconsin-Madison, Madison, WI, 53711, USA, Email: reynolds@astro.wisc.edu}
\author{S.I. Ipatov}
\affil{Univ. of Maryland, College Park, MD 20742, USA, Email: sipatov@umd.edu}
\author{A.S. Kutyrev}
\author{J.C. Mather}
\author{S.H. Moseley}
\affil{NASA - Goddard Space Flight Center, Greenbelt, MD, 20771, USA}

\newcommand{\btx}{\textsc{Bib}\TeX}

\maketitle

\begin{abstract}

We report on new observations of the motion of zodiacal dust using optical absorption line spectroscopy of zodiacal light.  We have measured the change in the profile shape of the scattered solar Mg I $\lambda$5184 line toward several lines of sight in the ecliptic plane as well as the ecliptic pole.  
The variation in line centroid and line width as a function of helio-ecliptic longitude show a clear prograde signature and suggest that significant fraction of the dust follows non-circular orbits that are not confined to the ecliptic plane. When combined with dynamical models, the data suggest that the zodiacal dust is largely cometary, rather than asteroidal, in origin. 

\end{abstract}

\vspace{-.1cm}
\section{INTRODUCTION}

The motion of interplanetary dust particles contains important information about their origin, distribution, and orbital evolution.  At optical wavelengths, dust with radii $\approx$\ 10-100$\mu$m that lie within $\approx$ 3 AU of the Sun scatters the incident solar radiation to produce zodiacal light, and the relative motion of the dust modifies the location and shape of solar spectral lines [1-3]. 
The fraction of zodiacal dust with a cometary or asteroidal origin is not well constrained at present [4-5], and the kinematics of these two components may shift the velocity and widths of the spectral features in unique ways. 
However, the low surface brightness of zodiacal light has, until recently, limited the observability of this effect, requiring a combination of high sensitivity and high spectral resolution.  
Previous work has shown the dust to be on prograde orbits, but different groups reported uncertain and contradictory results regarding the details of the orbital properties of the dust [6-7].
Here, we report on new measurements of scattered solar Mg I $\lambda$5184 absorption line in zodiacal light with the Wisconsin H-Alpha Mapper (WHAM), and compare the observations with predictions from dynamical models of the zodiacal dust cloud.

\section{OBSERVATIONS}
WHAM consists of a 15cm, dual-etalon Fabry-Perot spectrograph coupled to a 0.6m siderostat, and produces an average spectrum over a 1$^\circ$ circular field of view with a 12 km/s resolution within a 200 km/s spectral window. It is located at the Kitt Peak National Observatory in Arizona, and is entirely remotely operated. It is specifically designed to study extremely faint, diffuse optical light at high spectral resolution [8]. We have recorded spectra centered on the Mg I line at 5183.6\AA\ toward 49 directions along the ecliptic equator, with two directions at high ecliptic latitude on the nights of 2002 November 4 and 5 [9]. The unprecedented capabilities of WHAM allowed us to identify and remove several weak atmospheric emission lines that probably affected the results of previous investigations [6-7]. The terrestrial lines are stronger on the red side of the Mg I line. If they are not accounted for, they shift the line centroid to negative velocities and produce asymmetries in the line profile shape. 

A sample spectrum demonstrating the high resolution and sensitivity of the observations is shown in Figure 1. This spectrum was taken toward the north ecliptic pole, with the brightness of the line given in units of milli-Rayleighs per km/s.  The shaded solid line is a spectrum taken at twilight, where the light is scattered off the Earth's atmosphere which is at rest relative to the zodiacal cloud.  The two weak absorption lines in the twilight spectrum near +50 km/s are other solar lines of Fe I and Cr I.
The line centroid, width, and area of each spectrum were measured using a least-squares fitting technique, with the twilight spectrum used as a template. 
Figure 2 shows the change in velocity centroid with solar elongation (helio-ecliptic longitude, $\epsilon$) for directions along the ecliptic equator.  The data show a clear prograde kinematic signature. We see no evidence for a net radial outflow (at $\epsilon$=180$^\circ$) or an east/west asymmetry as reported by previous authors [6-7]. 

\begin{figure}[t]
\centering
\includegraphics[scale=0.4]{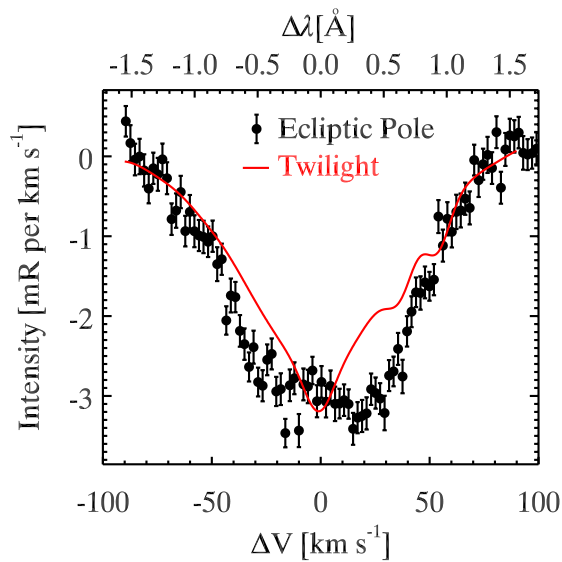}
\caption{\rm Spectrum of the zodiacal light taken toward the north ecliptic pole, centered on the Mg I $\lambda$5184 absorption line. The solid line shows the Mg I line from the twilight sky, and represents an unperturbed reference spectrum. Note the flat-bottomed profile and enhanced line width relative to the twilight spectrum.  This suggests that a significant population of the zodiacal dust particles are on non-circular orbits.}
\end{figure}

Figure 3 shows the change in the full-width at half-maximum of the lines as a function of elongation for directions along the ecliptic equator.  The solid horizontal line near 58 km/s is the width of the unperturbed twilight spectrum.  The Figure shows that along the ecliptic plane, the line is broadened by 10-20 km/s relative to the solar line, with no significant trends with elongation.
The large intrinsic width of the solar line makes these measurements difficult to quantify accurately and contributes to the large uncertainty compared to the velocity centroid data.

If all the dust particles were on pure circular orbits centered on the Sun, the zodiacal and twilight line profiles would be nearly identical toward the anti-solar direction. The broadening of the zodiacal profiles can only be explained by particles having radial components to their orbital motion. The data imply that a significant fraction of the dust follows elliptical orbits.

Furthermore, the substantial broadening of the profile toward the ecliptic pole (Figure 1), which is even greater than that toward the anti-solar direction (see [8]), implies a population of particles that have significant components of their orbital velocities projected perpendicular to the ecliptic plane. For an orbital speed of 30 km/s near 1 AU, and neglecting the radial motion associated with eccentricity, the approximately $\pm$15--20 km/s broadening at the base of the profile suggests a distribution of inclinations extending up to $\sim$30$^\circ$Ð-40$^\circ$ with respect to the ecliptic plane [9].

\begin{figure}[t]
\centering
\includegraphics[scale=0.4]{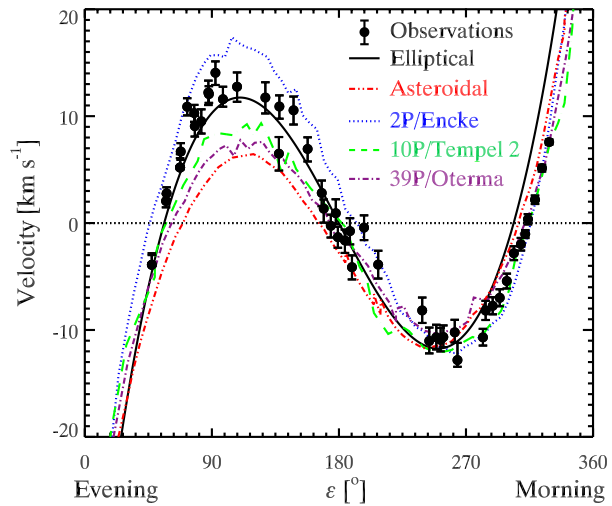}
\caption{\rm Velocity centroid of the Mg I absorption line as a function of helio-ecliptic longitude.  The observational data are shown as solid circles, with model predictions are shown as lines. The data indicate the dust follows prograde orbits, with no significant net radial outflow or east/west asymmetry. 
The observations are better fit by models in which the dust follows elliptical or cometary-like orbits (see text). }
\end{figure}

\section{COMPARISON TO MODELS}

The relationship between the observational data and the kinematic properties of the zodiacal dust cloud is a classical inversion problem.
The shape of the observed line profiles is determined by the population of dust particles of varying size, radial distance, scattering function, and relative motion along the line of sight.  
To infer the orbital properties of the particles that comprise the dust cloud, 
the data need to be compared to a range of predictive dynamical models. 
There has been considerable work done in the field of modeling the zodiacal dust cloud that take into account a wide range of observational characteristics [e.g., 10-12]. 
However, few models explicitly consider the dynamics of the cloud that can be directly compared to our data.
Below we discuss a few models from the literature, including some new models, that estimate the change in both the velocity centroid and line width with elongation.  

Some of the models that compare favorably to the data are shown in Figures 2 and 3. The solid black line is a fit to a model from [13].  This model describes particles on prograde, elliptical orbits with eccentricities uniformly distributed between 0 and 1, with randomly distributed perihelions. Their model did not include the influence of radiation pressure, and the particles were confined to the ecliptic plane. The model fits the centroid data well (Figure 2), but strongly overestimates the width of the lines (Figure 3).  The inclusion of radiation pressure and/or inclined orbits could provide a better match to the observations [9,14].

The colored lines in Figure 2 and 3 are models from [15-16], which trace the motion of different populations of dust particles subject to gravity, radiation pressure, and drag forces. The individual dashed and dotted lines represent particles with asteroidal and various cometary trajectories, with a ratio of radiation pressure to gravitational force of $\beta=0.002$.  
Some models considered particles with different $\beta$-values, and some with trans-Neptunian orbits, but those models are generally poorly matched to the data and are omitted for clarity.
In Figure 2, we see that a good match is provided by the cometary particles on inclined, eccentric orbits compared to the asteroidal particles.  In Figure 3, all of the models fall within the large scatter in the data, do not allow us to discriminate between the different models. Future observations of intrinsically narrower lines, such as Fe I,  will aid in using line widths to assess the asteroidal and/or cometary-type orbits of the zodiacal dust.

\begin{figure}[t]
\centering
\includegraphics[scale=0.4]{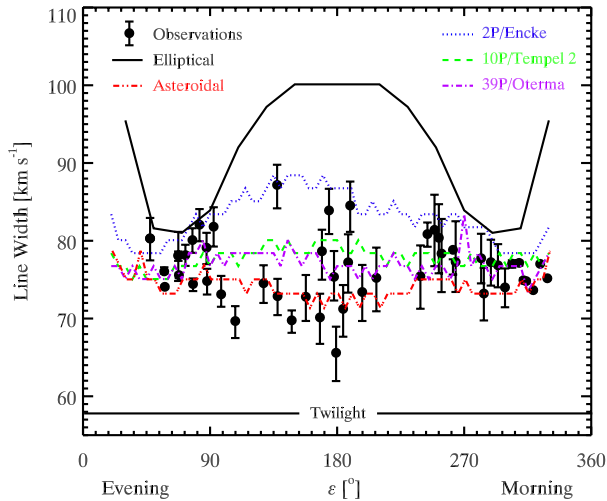}
\caption{\rm Full-width at half-maximum of the Mg I absorption line as a function of helio-ecliptic longitude.  The observations are shown as solid circles with the model data shown as various lines. The solid horizontal line represents the width of the absorption line taken at twilight.  The motion of the zodiacal dust broadens the line by $\approx$ 20 km/s relative to the twilight spectrum, with no significant changes in line width with elongation.  
With the exception of the model of [13], the models are generally well-matched to the observations, within the uncertainty.}
\end{figure}

\section{SUMMARY AND FUTURE WORK}

Observations of scattered solar absorption lines in the zodiacal light are a powerful technique for exploring the kinematics of the zodiacal dust cloud.  Our data are fit well by models that contain particles on elliptical orbits that are inclined to the ecliptic plane.  This suggests that most of the dust in the zodiacal cloud has a cometary origin, in agreement with [4-5].
The interpretation of our observations is highly model dependent, and we emphasize the importance of exploring a range of predictive models in assessing our conclusions.
Higher signal-to-noise observations covering a larger fraction of the ecliptic sky that include other, more intrinsically narrow, absorption lines will provide a more complete picture of the kinematics of the zodiacal dust cloud. New dynamical models that investigate a wider range of dust properties and orbital parameters can provide strong, quantifiable constraints on the nature of the zodiacal dust cloud when compared to these new observations. 

This research has been generously supported by the National Science Foundation through grant AST 02-04973 to the University of Wisconsin, and an MPS Distinguished International Research Fellowship grant AST 04-01416 to G.J.M.

\section{REFERENCES}

1. Gr\"{u}n E. et al. Dust in Interplanetary Space and in the Local Galactic Environment, in {\it{Astrophysics of Dust}}, ASP Conf. Series, 309, 245-264, 2004.  

2.\ Clarke D. et al. On the line profiles in the spectra of the zodiacal light, {\it{Astronomy \& Astrophys.}}, 308, 273-278, 1996. 

3.\ James, J. F. Theoretical Fraunhofer line profiles in the spectrum of the zodiacal light, {\it{MNRAS}}, 142, 45-52, 1969.

4. Dermott S.F. et al. Sources of Interplanetary Dust, in {\it{Physics, Chemistry, and Dynamics of Interplanetary Dust}}, ASP Conf. Series 104, 143-153, 1996.

5. Liou, J.-C., Dermott, S.F., Xu Y.L. The contribution of cometary dust to the zodiacal cloud. {\it{Planet. Space Sci.}} 43, 717, 1995.

6. Fried J.W. Doppler shifts in the zodiacal light spectrum, {\it{Astronomy \& Astrophys.}}, 68, 259-264, 1978. 

7. East I. R. \& Reay N. K. The motion of interplanetary dust particles. I - Radial velocity measurements on Fraunhofer line profiles in the Zodiacal Light spectrum, {\it{Astronomy \& Astrophys.}}, 139, 512-516, 1984.

8. Reynolds R.J. et al. The Wisconsin H-alpha Mapper (WHAM): A brief review of performance characteristics and early scientific results, {\it{Pub. Astr. Soc. Of Australia}}, 15, 14-18, 1998.

9. Reynolds R. J., Madsen G. J. \& Moseley, S. H. New Measurements of the Motion of the Zodiacal Dust, {\it{Astrophys. J.}}, 612, 1206-1213, 2004. 

10. Kelsall, T. et al. The COBE Diffuse Infrared Background Experiment Search for the Cosmic Infrared Background. II. Model of the Interplanetary Dust Cloud, {\it{Astrophys. J.}}, 508, 44-73, 1998.

11. Ozernoy, L.M. Physical modeling of the zodiacal dust cloud, in {\it{The Extragalactic Infrared Background and its Cosmological Implications}}, IAU Colloq. 204, eds. M. Harwit \& M.G. Hauser, 17-34, 2001.

12. Dikarev, V., et al. Upgrade of Meteoroid Model to Predict Fluxes on Spacecraft in the Solar System and Near Earth, in {\it{Dust in Planetary Systems}}, in press, 2006. 

13. Hirschi D.C. \& Beard D. B. Doppler shifts in zodiacal light, {\it{Planet. Space Sci.}}, 35, 1021-1027, 1987.

14. Rodriguez G. L. \& Magro C. S. The Doppler shift from Zodiacal Light, {\it{Astronomy \& Astrophys.}}, 64, 161, 1978.

15. Ipatov S. I. et al. Dynamical zodiacal cloud models constrained by high resolution spectroscopy of the zodiacal light, {\it{36th LPSC}}, (\#1266), 2005

16. Ipatov, S.I. \& Mather, J.C. Migration of Dust Particles to the Terrestrial Planets, in {\it{Dust in Planetary Systems}}, in press, 2006.

\end{document}